# Wideband Microstrip Antenna Design

**Scott A. Wilson, Graduate Student**
**Radar and Communications Lab**
**Department of Electrical Engineering**
**Penn State University, State College, PA**

**DESIGN GOAL**

Design a broadband microstrip antenna operating over the 450-700 MHz frequency range with a gain of at least 8 dB. The size should be minimized using appropriate choice of radiation element and substrate.

**DESIGN APPROACH**

A basic microstrip antenna (MSA) is typically low power, low gain (between 5-8 dB) and is fundamentally bandwidth limited to approximately 2 – 5% [1]. Design techniques have been developed in order to achieve higher bandwidths, as will be demonstrated in this paper. The aperture-coupled microstrip antenna (ACMSA) has been seen to achieve remarkably high bandwidths of nearly 70% with gain up to 8-9 dB [2]. The design of an ACMSA will be demonstrated as detailed in [3]. The following performance characteristics have been achieved:

| | |
|---|---|
| Center frequency | $f_c$ = 575 MHz |
| Antenna bandwidth | BW = 250 MHz |
| Percent bandwidth | BW (%) = 43% |
| Antenna gain | 8.43 dBi |
| Front-to-back radiation ratio | 14.1 dB |
| Efficiency | 99.984% |
| 3 dB beamwidth | ~ 80° |
| Overall dimensions | W = 22 cm, L = 21 cm, H = 10 cm |

I. Design Parameters

The ACMSA uses an aperture-coupled line feed to excite a resonant patch element. This idea was first proposed in 1985 as a bandwidth enhancement technique [3]. An exploded view of a single patch ACMSA can be seen in Figure 1. A variety of design parameters can be optimized for improved bandwidth and impedance matching.

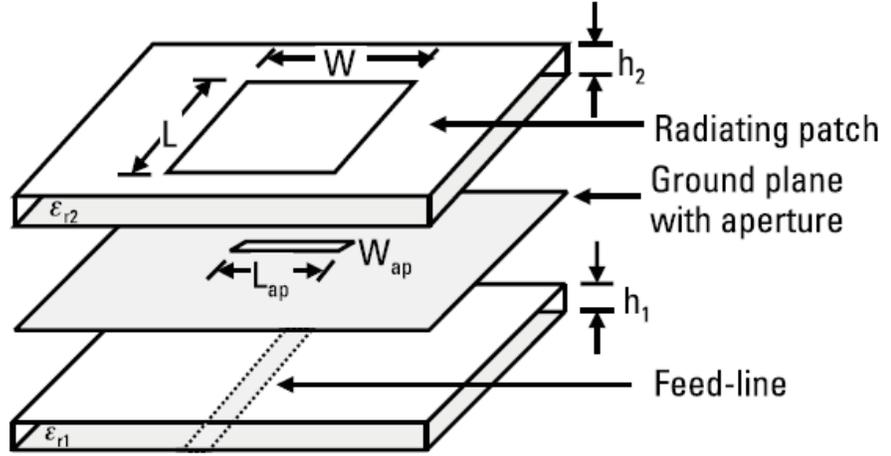

Figure 1: Exploded view of ACMSA [3].

A. Feed Line

Two major considerations for impedance matching are the feed line width and length. In order to achieve a characteristic impedance of 50 Ω, the feed line width needed to be determined for the particular substrate's thickness and dielectric constant. MATLAB was used to solve the roots of W for the following impedance equation:

$$Z_o = \frac{60}{\sqrt{\varepsilon_{eff}}} \ln(8\frac{h}{W} + \frac{W}{4h}) \qquad (1)$$

where $Z_o$ is taken to be 50 Ω, $h$ is the dielectric substrate thickness, and $\varepsilon_o < \varepsilon_{eff} < \varepsilon_r$ due to fringing effects. For thin substrates, $\varepsilon_{eff} \approx \varepsilon_r$.

The length of the feed line can be adjusted for impedance matching purposes. By extending the feed line, the reactance can be shifted to compensate for the capacitive coupling that occurs between the various layers. The impedance of the open-ended stub is given by

$$Z_s = -jZ_o/\tan(\beta L_s) \qquad (2)$$

where $Z_o$ is transmission line characteristic impedance and $L_s$ is stub length [3]. This parameter will inevitably need to be tuned and optimized after all design elements have been implemented.

B. Aperture Dimensions

The dimensions of the aperture play an important role in attaining high bandwidth. For a resonant slot ACMSA, the aperture length is set to be on the order of a half wavelength to achieve resonance at a particular frequency. The quality factor (Q) of this structure is predominantly influenced by the aperture width, as well as the spacing and dielectric permittivity of the substrate through which the feed line is coupled. To achieve low-Q (higher bandwidth), a thick substrate with relatively low dielectric permittivity should be used in order to reduce the amount of resonant coupling. A larger aperture will effectively provide a larger bandwidth, at the expense of a higher F/B ratio [3]. The amount of coupling that the aperture provides directly relates to the resistive part of the input impedance and therefore will also need to be tuned for optimal performance of the overall design.

C. Aperture Location

The location of the aperture with respect to the radiating patch element(s) will affect the input impedance to some degree. Slight misalignment in either x-direction or y-direction will not have a significant impact on the input impedance. Offsetting the aperture location may also induce asymmetries in the radiation pattern [3]. Therefore, the location of the aperture for this design will be centered for simplicity.

D. Patch Dimensions

For this design, two patches are used in order to provide an additional source of resonance. In general, the patch lengths should be on the order of a half wavelength for a given center frequency. The dimensions of the lower patch are very important as they influence the resonance between both the top patch as well as the aperture. The width of both patches can be increased in order to achieve higher bandwidth [3].

E. Substrates

For high bandwidth performance, each layer should generally be low permittivity with increased thickness. These properties have some of the greatest effects on bandwidth. For this design, a rigid Arlon DiClad substrate is used between the feed structure and slot plane, and thick foam layers with low permittivity ($\varepsilon_r \approx 1.07$) are used between the patches as seen in Figure 2.

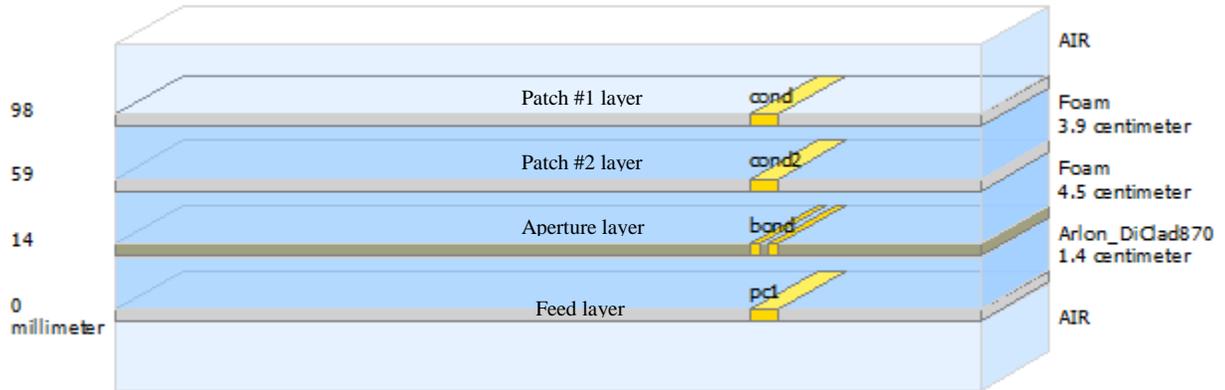

Figure 2: Substrate and copper stack-up for ACMSA design.

**DESIGN AND MODELING PROCEDURE**

To accomplish this design, Agilent's Advanced Design System (ADS) was used to model and simulate the ACMSA structure. Substrate properties and component geometries were roughly decided for 525 MHz operation. With reference to Figure 2, substrate layers were designed as such:

| Layer | Material | Permittivity | Loss Tangent | Thickness |
|---|---|---|---|---|
| Top Substrate Layer | Low-density foam | 1.07 | 0.0009 | 3.9 cm |
| Middle Substrate Layer | Low-density foam | 1.07 | 0.0009 | 4.5 cm |
| Bottom Substrate Layer | Arlon DiClad870 | 2.33 | 0.0013 | 1.4 cm |

These low-permittivity, high thickness substrates are a good starting point for building low-Q resonant structures. The final dimensions for the feed structure, slot and patch elements are detailed in Figures 3a – 3d. The superimposed 2D stack-up of each layer is seen in Figure 4, and a 3D stack-up view is shown in Figure 5. These dimensions were decided based off of the discussion in the previous section. Parametric tuning was performed manually. However, ADS provides parametric optimization tools for achieving optimum system performance.

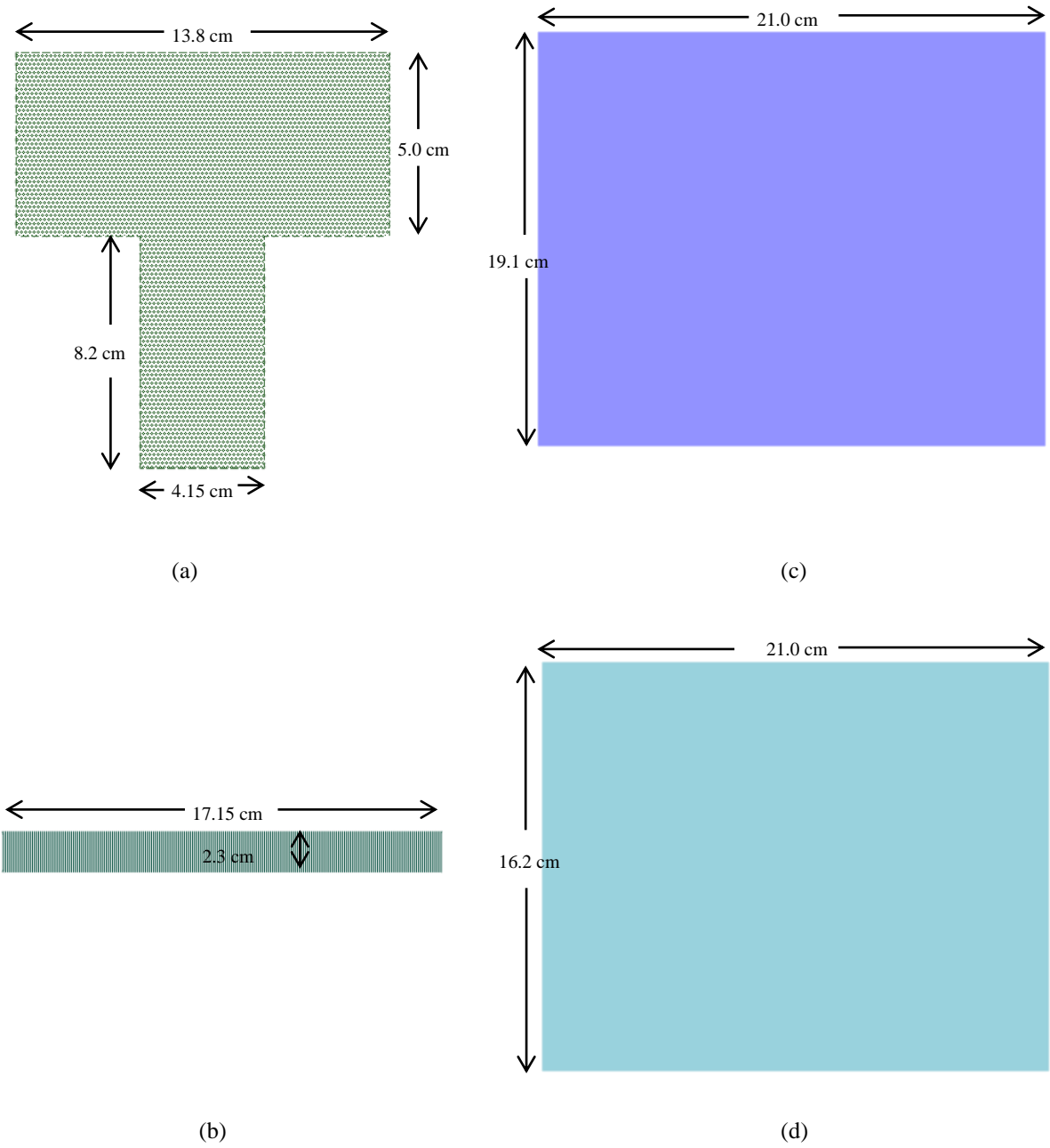

Figure 3: Physical dimensions for (a) feed line structure, (b) aperture slot, (c) middle patch, and (d) top patch.

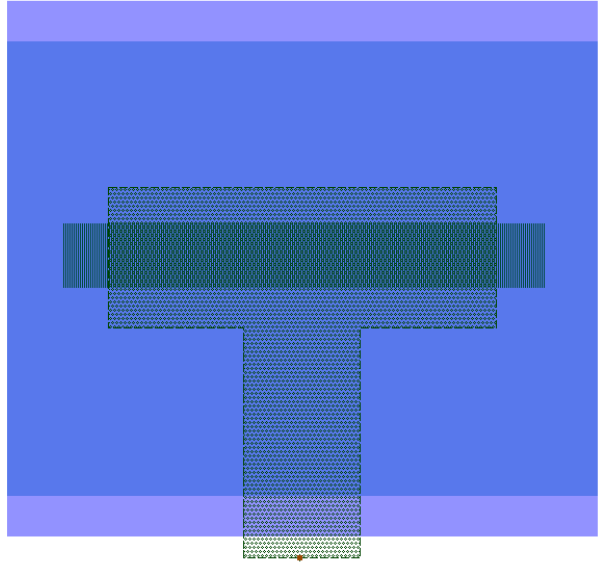

Figure 4: 2D stack-up of ACMSA structure.

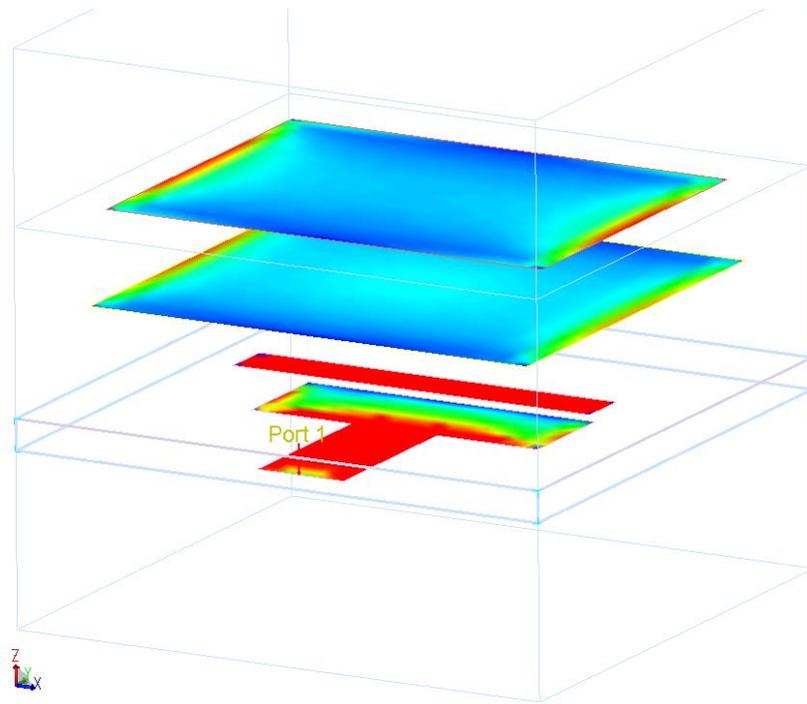

Figure 5: 3D stack-up and electric field distribution for ACMSA structure.

**DISCUSSION OF RESULTS**

Using the techniques previously discussed for aperture-coupled antenna design, a bandwidth of 43% was attained with a relatively balanced response over the passband. Phase response is also relatively linear, providing good wideband frequency characteristics.

Resonant coupling between the feed structure, slot and middle patch is responsible for the first minimum seen in the S11 plot of Figure 6. The Q of this valley can be adjusted by varying the substrate thicknesses, slot width or patch width. Each of these parameters was manually tuned to obtain lower resonant coupling and a lower center frequency with respect to the top patch resonant frequency. This was done in order to extend the lower frequency response. Optimization tools can be used to further optimize these parameters. Similarly, the upper minimum results from resonant coupling between the two patch elements. The top patch dimensions were adjusted to achieve resonance just below 700 MHz, while the middle patch dimensions were manually tuned to achieve higher bandwidth over both resonant structures.

The impedance plot seen in the Smith chart of Figure 7 also indicates good bandwidth response by the tight loops of the center loci. When the impedance is roughly matched in both real and imaginary components, antenna radiation and high efficiency are achieved. The double loop is a result of the multiple resonant structures, providing wider bandwidth.

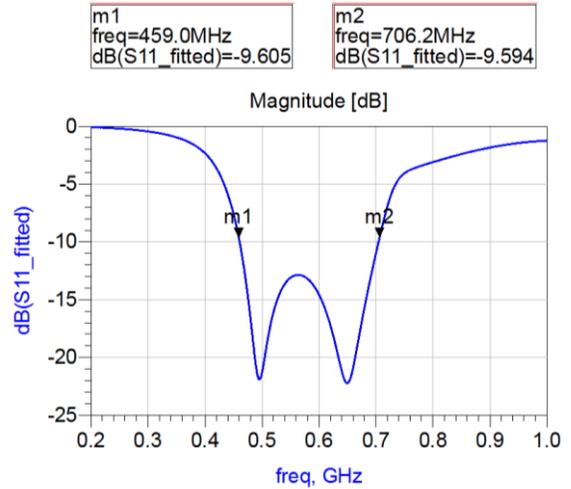

Figure 6: S11 magnitude response.

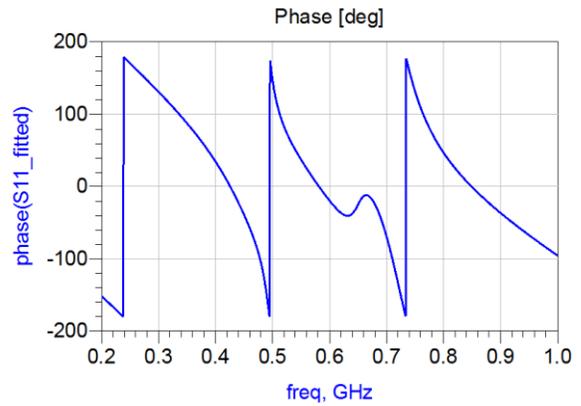

Figure 7: S11 phase response.

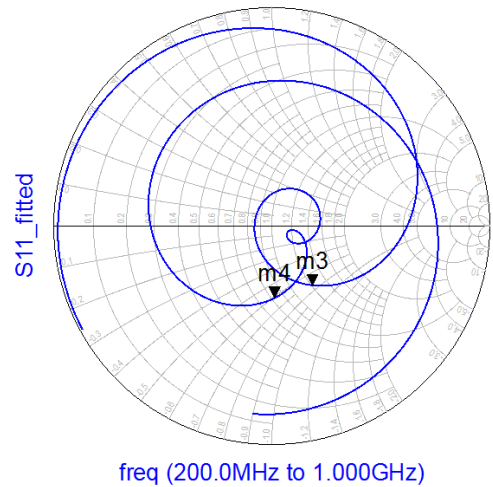

Figure 8: Impedance vs. frequency.

The far-field radiation pattern seen in Figure 9 reveals a fair amount of back-end radiation. This effect could be suppressed by reducing the aperture width, or perhaps by adding another substrate backing layer. The F/B ratio achieved of 14.1 dB is around the same level as was achieved in [2]. The 2D azimuthal cuts for the radiation pattern are seen in Figure 10. The antenna beamwidth is approximately 80° and the cross-polarization response is nearly 60 dB down from the co-polarization response at boresight angle. Additional antenna parameters are reflected in Figure 11.

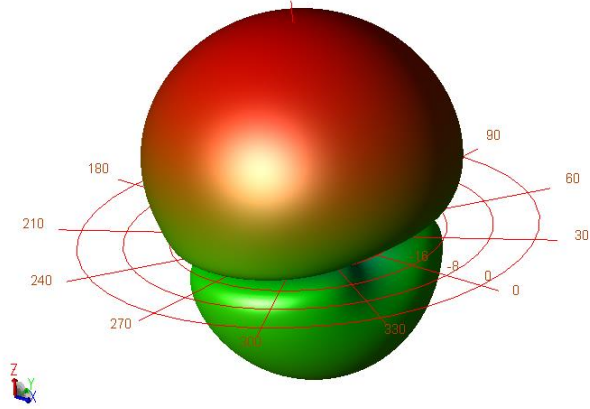

Figure 9: 3D far-field radiation pattern for ACMSA.

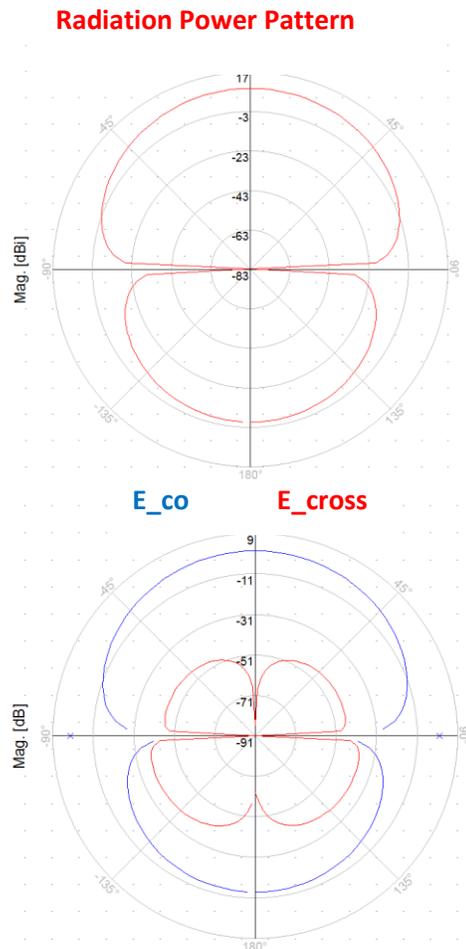

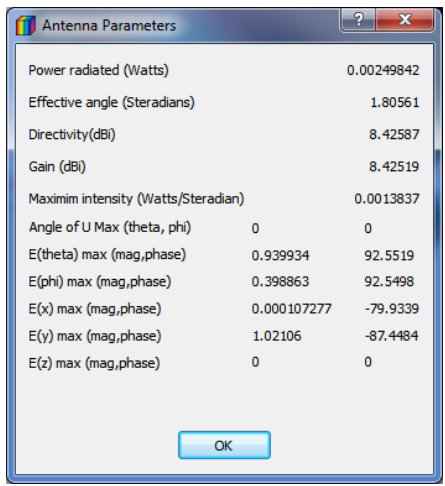

Figure 11: Antenna performance parameters.

Figure 10: Radiation and polarization patterns.

## CONCLUSIONS

Through modeling and simulation, a wideband microstrip patch antenna was achieved. This design could be further optimized using ADS's optimization tools. Given more time, parameterization of each component dimension could be used to achieve better results. This optimization could be done by defining performance goals, parametric variables, and various constraints in order to achieve an optimal design. Different substrate thicknesses and permittivities could also be evaluated. Furthermore, additional resonant structures could be added, with the trade-off of greater design complexity. Overall, good antenna performance was achieved using a double-stacked aperture-couple microstrip antenna structure.